\documentclass[referee]{mn2e}
\usepackage{graphicx}

\title[The observed infall of galaxies towards the Virgo cluster]
{The observed infall of galaxies towards the Virgo cluster}

\author[I.\,D.\,Karachentsev, O.\,G.\,Nasonova (Kashibadze)]
{I.\,D.\,Karachentsev$^{1}$\thanks{E-mail:ikar@sao.ru},
O.\,G.\,Nasonova (Kashibadze)$^{1}$\thanks{E-mail:phiruzi@gmail.com}\\
$^{1}$Special Astrophysical Observatory of the Russian Academy of
Sciences, Nizhnij Arkhyz, KChR, 369167, Russia}

\begin{document}


\pagerange{\pageref{firstpage}--\pageref{lastpage}} \pubyear{2010}

\def\LaTeX{L\kern-.36em\raise.3ex\hbox{a}\kern-.15em
    T\kern-.1667em\lower.7ex\hbox{E}\kern-.125emX}

\newtheorem{theorem}{Theorem}[section]

\label{firstpage}

\maketitle

\begin{abstract}
We examine the velocity field of galaxies around the Virgo cluster
induced by its overdensity. A sample of 1792 galaxies with distances
from the Tip of the Red Giant Branch, the Cepheid luminosity, the SNIa
luminosity, the surface brightness fluctuation method, and the
Tully-Fisher relation has been used to study the \textit{velocity-distance}
relation in the Virgocentric coordinates. Attention was paid to some
observational biases affected the Hubble flow around Virgo.

We estimate the radius of the zero-velocity surface for the Virgo cluster
to be within (5.0 -- 7.5) Mpc corresponding to
$(17 - 26)^{\circ}$ at the mean cluster distance of 17.0 Mpc.
In the case of spherical symmetry with cosmological parameter
$\Omega_m=0.24$ and the age of the Universe $T_0= 13.7 Gyr$, it yields
the total mass of the Virgo cluster to be within
$M_T=(2.7 - 8.9)\cdot10^{14}M_{\sun}$ in reasonable
agreement with the existing virial mass estimates for the cluster.
\end{abstract}

\begin{keywords}
galaxies: clusters: individual: Virgo -- dark matter.
\end{keywords}

\section{Introduction}

The gravitational action of mass of a solitary system of galaxies leads
to deceleration of the local Hubble flow. As a result, the line of
average velocity of neighboring galaxies relative to the center of a
cluster (or a group) deviates from the linear Hubble relation, going to
negative values at small distances $R < R_0$. Here, $R_0$ means the
radius of the zero-velocity surface, which separates the galaxy system
against the global cosmic expansion. As it was shown by Lynden-Bell (1981)
and Sandage (1986), in the simplest case of spherical symmetry with
cosmological parameter $\Lambda=0$ the radius $R_0$ depends only on
the total mass of a group $M_T$ and the age of the Universe $T_0$:

$$M_T=(\pi^2/8G)\cdot R_0^3\cdot T^{-2}_0,                   \eqno(1) $$
where $G$ is the gravitational constant. Measuring $R_0$ via distances
and radial velocities of galaxies outside the virial radius of the system
$R_{vir}$, one can determine the total mass of the system independent
of its virial mass estimate. Note that both the methods deriving
mass from internal and from external galaxy motions correspond to
different linear scales where $R_0$ is roughly 4 times as large as the
virial radius. In reality, galaxy groups and clusters do not have perfect
spherical symmetry, and cosmology with $\Lambda=0$ is not true.

Numerous measurements of distances to nearby galaxies obtained recently
with the Hubble Space Telescope (HST) allowed us to investigate the
Hubble flow around the Local Group and other proximate groups
(Karachentsev et al. 2002, Karachentsev \& Kashibadze 2006, Karachentsev
et al. 2006, 2009). The radii $R_0$ obtained from observations for
nearby groups around: the Milky Way \& Andromeda (the Local Group), M81,
Centaurus~A, Maffei \& IC~342, NGC~253 (Sculptor filament), and NGC~4736
(Canes Venatici I cloud) are ranged within ($0.7 - 1.4$~Mpc). The average ratio
of total-to-virial masses for these six groups, derived from $R_0$ via eq.~(1) and from
$R_{vir}$, turns out to be $<M_T/M_{vir}> = 0.60\pm0.15$ (Karachentsev, 2005).
But as it was noticed by Peirani \& Pacheco (2006, 2008) and Karachentsev
et al. (2007), in
a flat universe dominated by dark energy the resulting $M_T(R_0)$ mass is
higher than that derived from the canonical Lema\^\i{}tre-Tolman eq.~(1). In
the "concordant" cosmological model with $\Lambda$-term and $\Omega_m$ as a
matter component eq.~(1) takes a form
$$
 M_T =(\pi^2/8G)\cdot R_0^3\cdot H^2_0/f^2(\Omega_m), \eqno(2)$$
\noindent{}where
$$
f(\Omega_m)=(1-\Omega_m)^{-1}-(\Omega_m/2)\cdot(1-\Omega_m)^{-3/2}\cdot
arccos h[(2/\Omega_m)-1].\eqno(3)
$$
\noindent{}Assuming $\Omega_m=0.24$ and $H_0=72$ km s$^{-1}$ Mpc$^{-1}$
that corresponds to $T_0=13.7$ Gyr (Spergel et al. 2007), one can rewrite
(2) as

$$(M_T/M_{\sun})_{0.24}=2.12\cdot10^{12}(R_0/Mpc)^3.         \eqno(4)$$

It yields the mass that is 1.5 as large as derived from the classic eq.~(1).
This correction leads to a good agreement in average between the $R_0$
mass estimates and virial masses for the  abovementioned galaxy groups.

For galaxies around the nearest cluster in Virgo, expected velocity
deviations from the pure Hubble flow (so-called Virgocentric infall)
were regarded in dynamical models by Hoffman et al. (1980), Tonry \&
Davis (1981) and Hoffman \& Salpeter (1982). These authors note that
with the virial mass of the Virgo cluster
$M_{vir}\sim6\cdot10^{14}M_{\sun}$, the radius of the zero-velocity
surface around the cluster amounts to $\sim27^{\circ}$, i.e. the infall
zone covers nearly 1 steradian of the sky. According to Hoffman et al.
(1980), the observed decrease of radial velocity dispersion within the
angular distance $\Theta=[0 - 24]^{\circ}$ from the Virgo center for 228
galaxies agrees, in the main, with the Virgocentric infall pattern for
the cluster mass mentioned above. Tully \& Shaya (1984) considered the
phenomena of infall of galaxies towards Virgo both in the point-mass and
distributed-mass models for the cluster with different values of the
cosmological parameter $\Omega_{\Lambda}$ and the age of the Universe
$T_0$. Using Tully-Fisher distance estimates for 19 galaxies inside the
virial radius of $\sim6^{\circ}$ and 14 galaxies outside it, the authors
ascertain the expected infall with $R_0\sim28^{\circ}$.

Later, Tonry et al. (2000, 2001) developed a model of the Virgocentric
flow basing on accurate distance measurements for 300 E and S0 galaxies
from surface brightness fluctuations. Their model fits well the
observational data on galaxy distances and radial velocities for the
Virgo cluster distance of 17.0 Mpc and its total mass of
$7\cdot10^{14}M_{\sun}$. According to their model, our Local Group has a
peculiar velocity of 139 km s$^{-1}$ directed towards the Virgo center.

Teerikorpi et al. (1992), Ekholm et al. (1999) and Ekholm et al. (2000)
examined the Virgocentric flow with different models of density
distribution in the cluster and infered expected relations between
velocities and distances of galaxies relative to the cluster center.
Using Cepheid distances to 23 galaxies and Tully-Fisher distances to 96
galaxies, the authors conclude that the radius of the zero-velocity
surface ranges from 20$^{\circ}$ to 31$^{\circ}$, and the total cluster
mass is equal to $(1\div2)$ of its virial value.

During the last decade, the observational database on distances to
galaxies in a wide vicinity of the Virgo cluster has grown significantly,
allowing to determine $R_0$ and, therefore, the total mass of the Virgo
cluster with better accuracy.

\section{The observational samples}

To examine phenomena of the Virgocentric flow, we used distance
moduli of galaxies from different publications preferring more precise
measurements. The main data sources are listed below.

A) Taking luminosity of the Tip of the Red Giant Branch (TRGB) as a
standard is the most efficient and the most universal method to
determine distances to nearby galaxies, as it is practically independent
of their morphological type. Being applied to galaxy images in two ore
more photometric bands obtained with WFPC2 or ACS cameras at the HST,
the TRGB method yields an accuracy of distance measurements of $\sim7$\%,
as it was founded by Rizzi et al. (2007). A consolidated list of distances
for the Local Volume galaxies is presented in the Catalog of
Neighboring Galaxies (= CNG, Karachentsev et al. 2004). The CNG sample
of 451 galaxies has been collected based on two conditions: galaxy
distance $D < 10$~Mpc, if a galaxy has individual distance estimate; otherwise
galaxy radial velocity with respect to the Local Group $V_{LG} < 550$~km~s$^{-1}$.
Below we use from CNG the galaxies with only TRGB or Cepheid distances,
supplying them with new TRGB distances from recent publications
(Karachentsev et al. 2006, Tully et al. 2006).

B) The surface brightness fluctuation method, SBF, applying to early
type galaxies, assumes that the old stellar population (RGB) is
prevailing in a total luminosity, and the galaxy structure does not
suffer from irregularities due to dust clouds. Using this approach,
Tonry et al. (2001) determined SBF distances to 300 E and S0 galaxies
with a typical errors of $\sim12$\%. This sample is distributed over the
whole sky extending to cz $\sim 4000$ km~s$^{-1}$ with a median
velocity of 1800 km~s$^{-1}$.

C) Mei et al.(2007) undertook a two-color ACS/HST imaging survey for 100
early-type galaxies situated in the Virgo cluster core (the ACSVCS project).
They derived precise SBF distances to 84 E, S0 galaxies with a typical
error of 8\%, and revealed a 3D-shape of the Virgo cluster to be a slightly
triaxial ellipsoid with axis ratios of (1:0.7:0.5). We expanded the ACSVCS
sample with other precise SBF and TRGB distance measurements in the Virgo
core made by Neilsen \& Tsvetanov (2000) and Caldwell (2006) with ACS/HST
and by Jerjen et al. (2004) with the VLT. That yields us the total "ACSVCS+"
sample of 116 galaxies.

D) In wide vicinity of the Virgo cluster there are 22 galaxies with
distances measured by Tonry et al. (2003) via SNIa. This sample is
small but has the distance error of only 5\%.

E) Based on the "2MASS Selected Flat Galaxy Catalog" (Mitronova et al.
2004 = 2MFGC), Kashibadze (2008) determined distances to 402 spiral
edge-on galaxies with radial velocities $<3000$ km s$^{-1}$. A
multiparametric NIR Tully-Fisher relation was applied to them, yielding
a typical distance error of $\sim20$\%. The zero point of the
luminosity - line width relation was calibrated by 15 galaxies with
Cepheid and TRGB distance measures.

F) Finally, the former samples of galaxies were supplemented with a
compilation of distances by Tully et al. (2008, 2009) that have been
obtained from optical (B, R or I band) one-parametric Tully-Fisher
relations. This compilation relays on numerous HI-line and photometric
observations carried out by Methewson \& Ford (1996), Haynes et al.(1999),
Tully \& Pierce (2000), Koribalski et al. (2004), Springob et al. (2005),
Theureau et al. (2006) and other authors. Zero points of the data were
recalibrated by a set of 40 galaxies with known Cepheid and TRGB
distances. As a last step, we used also distances from a very big and
important SFI++ sample (Springob et al. 2007), which have not been included
in Tully et al. (2009) compilation. At total, we used distance estimates
for 941 spiral galaxies whose radial velocities were limited by
3000 km~s$^{-1}$. The typical distance error for them is $\sim20$\%,
though there are some cases with much higher errors due to uncertainties
of galaxy inclination, presence of interacting companions or HI profiles
of low quality.

A substantial overlap between the two last {\itshape luminosity-linewidth}
samples provides confirmation that their zero points are the same
and gives rms agreement per measure of 0.40 mag. As it is seen from
Fig.\,1 by Tully et al. (2008), there is an excellent agreement in
distance moduli between the {\itshape luminosity-linewidth} and other
(Cepheid, TRGB, SBF and SNIa) measures. In particular, for 12 galaxies
in our list with both TF and TRGB or SNIa moduli the mean distance
difference is ($-1.4 \pm 1.2$)~Mpc, while for 20 galaxies with SBF or
TRGB moduli the average difference is only ($0.2 \pm 0.2$)~Mpc.

Following the previous authors (Tully \& Shaya 1984, Ekholm et al. 2000)
we have formed a composite sample of galaxies limiting their angular
separation from the Virgo center to  $\Theta<30^{\circ}$. We have
considered the radiogalaxy Virgo~A = NGC~4486 to be the physical center
of the cluster as its position is close to the center of X-ray emitting
gas. The total number of galaxies in this cone volume with apex angle
$\Theta<30^{\circ}$ is 630 (the {\itshape 2D sample}).

Limiting the angular separation of galaxies from the Virgo center
introduces some selection effects into the Virgocentric flow analysis.
For this reason we have used also another way to form the observational
sample, considering galaxies with spatial distances from the Virgo
cluster center $R_{vc}<30$ Mpc (the {\itshape 3D sample}). This approach
suffers a drawback too because galaxy distances are measured with errors
and their significance is different at the proximate and the distant
boundary of the spherical volume (the so-called \textit{Malmquist bias}).
The total number of galaxies in our 3D sample amounts to 1792, and the
fractions of diverse subsamples differ significantly from that in the 2D
sample.

Table 1 presents the summary of observational data that we have used.
The first column indicates subsample kinds, the second one gives typical
distance errors expressed in magnitudes. Columns (3) and (5) contain
numbers of galaxies in the cone (2D) or in the spherical (3D) volume.
The {\itshape sample goodness} $G$, defined as
$G=(N/100)^{1/2}\cdot\sigma^{-1}_m$, is a useful parameter which
characterizes a statistical weight of a certain sample (Kudrya et al.
2003). Goodness values are indicated in columns (4) and (6). For example,
the subsample of galaxies with SNIa distances is scanty but its statistical
significance is comparable with that of other samples because of higher
accuracy of distance measurements. As one can see, the galaxy subsample
ACSVCS+ has the maximum statistical weight in the 2D set, however almost all
these galaxies are concentrated within the virial radius. In the 3D set
the highest goodness corresponds to the TRGB sample, but its majority
is crowded on the nearby side of the examined volume. The last two TF
samples exhibit a significant increase in number going from the 2D to
the 3D samples that is caused by the well-known effect of morphological
segregation of late-type vs. early-type galaxies along the cluster radius.

\section{Radius of the zero-velocity surface $R_0$}

The virial radius of the Virgo cluster $R_{vir}=1.8$ Mpc (Hoffman et al.
1980) corresponds to its angular scale of 6.0$^{\circ}$,
assuming the average distance to the cluster members to be
17.0 Mpc. Radial velocities and distances relative to the Local Group
(LG) centroid for 259 galaxies in this zone are represented in the top
panel of Fig.\,1. Here, precise distances for most the galaxies were
obtained within the special survey ACSVCS with HST (Mei et al. 2007).
The Virgo cluster members, located in the distance range from 14 to
20 Mpc, demonstrate a radial velocity scatter from $-$800 up to
$+$2300 km s$^{-1}$. Foreground galaxies are scarcely presented on the
panel while background objects tend to lie below the linear Hubble
regression with the global Hubble parameter
$H_0=72$ km s$^{-1}$ Mpc$^{-1}$ (Spergel et al. 2007),
showing thereby the expected effect of infall into the Virgo cluster
from the opposite side. The centroid of galaxies forming the "virial
column" at [$17.0\pm1.8$] Mpc, marked by vertical
lines, has a mean velocity $+$1004$\pm$70 km s$^{-1}$ versus the
expected Hubble velocity of $+$1224 km s$^{-1}$ at the distance of 17.0
Mpc, which can be explained by a peculiar motion of the Local Group
$\sim220\pm70$ km~s$^{-1}$ directed towards Virgo. The dotted and solid S-shaped
curves correspond to a Hubble flow perturbed by a point-like mass of
$2.7\cdot10^{14}M_{\sun}$ and $8.9\cdot10^{14}M_{\sun}$ (as the limiting
cases discussed below) for the line-of-sight passing exactly through
the cluster center.

The distributions of radial velocities and distances for remaining
galaxies of the 2D sample in close surroundings of Virgo,
($6^{\circ} < \Theta<15^{\circ}$) and in a distant periphery
($15^{\circ} < \Theta<30^{\circ}$), are shown in the middle and the
bottom panels of Fig.\,1. Here, the solid and dotted S-shaped lines
having lower amplitudes describe the behavior of perturbed Hubble flow
at angular distances $\Theta$ equal $6^{\circ}$ and $15^{\circ}$,
respectively. These panels display some signs of the infall effect
too, however, in front of the Virgo, the expected infall is seen barely.

Considering a set of such kind Hubble diagrams with their different
amplitudes of S-shaped waves decreasing with the angular distance $\Theta$,
one can find the quantity of the cluster mass that fits the observed
infall pattern in the best way. But this approach seems us to be not
transperent enough. In order to determine $R_0$ and the total cluster
mass via it, we have converted our observational data into distances
and velocities expressed relative to the cluster center.
The top panel of Fig.\,2 shows the layout of a galaxy ($G$)
relative to the observer ($LG$) and the cluster center ($C$)
with angular separation $\Theta$ from the cluster center. The spatial
distance of the galaxy from the center therefore is

$$R^2_{vc}=R^2_g+R^2_c-2R_gR_c\cos\Theta.                    \eqno(5)$$

Assuming that the galaxy and the cluster center are involved in an
almost unperturbed Hubble flow ({\itshape Hf} case) with negligible
peculiar velocities, we can state the mutual velocity difference between
$G$ and $C$ in projection onto the stright line connecting them as

$$V_{vc}=V_g \cos{\lambda}-V_c \cos\mu,                       \eqno(6)$$
where $\mu=\lambda+\Theta$ and
$\tan\lambda= R_c \sin\Theta/(R_g-R_c \cos\Theta)$.

The distribution of galaxies in the Virgocentric reference frame
$\{V_{vc}, R_{vc}\}$ is represented in Fig.\,3. The only 391 galaxies
with angles $\lambda$ obeying $\lambda<45^{\circ}$ or
$\lambda>135^{\circ}$ from the whole 2D sample are shown here. Selecting
galaxies situated approximately in front and behind the cluster is meant
to reduce the role of tangential velocity componens which remain still
unknown. The polygon curve traces the running median with a window of 1
Mpc. The median follows roughly the linear Hubble regression with
$H_0=72$ km s$^{-1}$ Mpc$^{-1}$ (the inclined dashed line) at middle
Virgocentric distances of $\sim$15 Mpc, but tends to deviate from the
$H_0 \cdot R_{vc}$ line at smaller scales, crossing the zero-velocity line
at $R_0\simeq6$ Mpc. The behaviour of the running median at large scales is
strongly skewed by a selection effect due to the adopted limit for
galaxy velocities $V_{LG} <3000$ km s$^{-1}$.

Another approach can be used also for converting the observational
radial velocity of a galaxy $V_g$ into its Virgocentric velocity (the
case of pure Virgocentric flow, {\itshape Vf}). If the galaxy is not
involved in the general cosmological expansion but is falling instead
towards the Virgo cluster with a velocity $V_{in}$ ( Fig.\,2b), than its
radial velocity relative to the observer will be expressed as

$$ V_g=V_c\cdot\cos\Theta-V_{in}\cos\lambda,                  \eqno(7)$$
and the infall velocity $V_{in}$ itself can be written as

$$V_{in}=(V_c \cos\Theta-V_g) \sec\lambda,                    \eqno(8)$$
where the angles $\Theta$ and $\lambda$ are shown in the Fig.\,2b.
Evidently, the discrepancy between these two extreme approaches,
{\itshape Hf} and {\itshape Vf}, decreases when $\lambda$ tends to 0
or to 180$^{\circ}$.

As can be easily seen, selecting galaxies both in the cone with apex angle
$\Theta=30^{\circ}$ and the angle $\lambda$ entails a loss of galaxies
at large Virgocentric distances which becomes apparent in the top right
corner of the Fig.\,3. A selection of galaxies in the volume
$R_{vc}<30$ Mpc (3D sample) reduces the bias appreciably. Fig.\,4
represents the sky distribution of galaxies with known Virgocentric
distances up to 30 Mpc in equatorial coordinates. The galaxies of this
3D sample are marked as circles and their diameters indicate three
distance ranges: (0--12) Mpc, (12--22) Mpc, and more than 22 Mpc from
the observer. This map exhibits that the selection of galaxies by
their angle $\Theta<30^{\circ}$ brings some systematic skews dependent
on distance. In particular foreground Virgo galaxies are preferentially
removed by this angular selection.

The Hubble diagram for 1792 galaxies of the 3D sample is represented in
Fig.\,5. The symbols for objects from different sources of distance data
are the same as for Fig.\,1. The number of galaxies in the 3D sample
is roughly 3 times as large as in the 2D sample. It is noteworthy, they
populate the crucial regions in front and behind the Virgo cluster more
thoroughly giving us an opportunity for more detailed analyses of the
Virgocentric infall. The velocity - distance relation for these galaxies
with respect to the cluster center is shown in Fig.\,6. Its top panel
corresponds to the assumption of pure Hubble flow of galaxies
({\itshape Hf} case) while the bottom panel represents the case of pure
radial motions towards the Virgo center ({\itshape Vf}). As previously,
the galaxies located far away from the line of sight crossing the
cluster center, i.e. with 45$^{\circ}<\lambda<135^{\circ}$, are
eliminated in order to reduce the role of unknown tangential velocity
components. This condition diminishes the number of sampled galaxies by
42\%.

Comparison the bottom panel of Fig.\,6 with the top one shows that
switching from the {\itshape Hf} case to the {\itshape Vf} case does
not lead to any dramatic changes in the Hubble flow pattern given in the
Virgocentric coordinates. Some galaxies move along the vertical axis
appreciably but the total behaviour of the running medians traces
the infall of galaxies towards the cluster in a similar way. The
asymptotic tendency of the median at large distances $R_{vc}$ looks
much more regular for the 3D sample than for the 2D one.

Our elimination of galaxies with 45$^{\circ}<\lambda<135^{\circ}$ is
a little arbitrary. To estimate the response of the
Virgocentric flow pattern to changing this condition we have imposed
also a more rigid constraint, eliminating galaxies with
$30^{\circ}<\lambda<150^{\circ}$. The corresponding diagrams for
{\itshape Hf} and {\itshape Vf} cases are represented in two panels of
Fig.\,7. As it is seen, the more severe selection of galaxies by $\lambda$
reduces their number by more than a half. However, the behaviour of the
running median is almost the same.

In the bottom panel of Fig.\,7 every galaxy from the samples A,B,C, and D
(see Table 1) having a distance error within 12\% is supplied by error
bars indicating where the galaxy should be situated if its distance
from the observer changes by $\pm1\sigma_D$. As expected, these error
bars are much longer for galaxies situated behind the cluster.
Changing a galaxy distance by $\pm12$\% leads in some regions of the
\{$V_{vc},R_{vc}$\} diagram to significant displacement of the galaxy
along the both Virgocentric coordinates, and therefore to appreciable
galaxy skips relative to the zero-velocity line. (We do not include
the longer error bars for the Tully-Fisher distances).

To quantify the uncertainties on the running median curves plotted
in Figs. 6 and 7, we generated extensive sets of bootstrap realisations.
Their results permit us to estimate the radius of the zero velocity
surface and also its rms error presented in Table 2. The table columns
contain: (1) size of the window for a running median, taken to be 0.8,
1.0, and 1.2 Mpc; (2,3) the mean radius of the zero velocity surface
$R_0$ and its rms scatter obtained with regarding to the assumptions
on the pure Hubble flow ({\itshape Hf}) and the pure Virgocentric
flow ({\itshape Vf}), respectively; here only four samples: A,B,C, and D
with precise distance moduli were taken into account; (4,5) the same
quantities for the samples E and F with Tully-Fisher distances; (6,7)
the radii $R_0$ and their errors for the total set of available data
on galaxy distances. We shell discuss the interpretation of these
different estimates of $R_0$ in the next section.

\section{Discussion and conclusions}

We have analysed available observational data on distances and radial
velocities of galaxies in wide surroundings of the Virgo cluster in
order to study the Virgocentric infall. The main purpose of the present
paper is to determine the radius of the zero-velocity surface $R_0$
that separates the cluster against the global cosmic expansion. By using
this observational quantity we are able to estimate the total mass of the
Virgo concentrated within $R_0$ and compare it with the virial mass
estimates corresponding to roughly 4 times lower scale, $R_{vir}$.
Based on the Hubble diagrams transformed into Virgocentric coordinates
and the results of our bootstrap numerical experiments given in Table 2,
we derive $R_0$ to be in the range of $5.0 - 7.5$~Mpc with a typical random
error of 1.0~Mpc. The derived value of $R_0$ can be affected by some
systematic circumstances.

  Analysis of the Hubble diagrams in the Virgocentric coordinates
suffers the lack of data on tangential velocity components of the
galaxies. We tried to overcome this drawback  with assumptions
regarding a dominant type of galaxy motions in the Virgo proximity.
Conversion of the observed radial velocities of galaxies into their
Virgocentric velocities was performed by us under two extreme kinematic
assumptions: almost unperturbed Hubble flow ({\itshape Hf}) or
almost pure radial flow towards Virgo ({\itshape Vf}).
We found that adopting one or another scheme doesn't change
significantly a general pattern of the Virgocentric infall. Calculating
velocities in the {\itshape Vf} case yields, at the average, some
smaller values of $V_{vc}$, which causes a slightly larger
($+$0.7 Mpc) value of $R_0$.

The $R_0$ quantities presented in Table 2 tend to be correlated
with the smoothing window size. When the window changes from 0.8~Mpc
to 1.2~Mpc, the radius dicreases on 0.3~Mpc at average. Variation
the window size in wider range: from 0.5~Mpc to 2.0~Mpc still
leaves $R_0$ within its random errors.

   The most noticeable systematic variations in the radius $R_0$ are
seen in dependence on galaxy samples. The use of samples of galaxies
with only precise distance measurements (via Cepheids, TRGB, SBF, and SNIa)
yields the radius $R_0$ within  $6.5 - 7.5$~Mpc, while the use of less
precise Tully-Fisher distances leads to $R_0$ estimates around 5.2~Mpc.
The physical origin of this difference is clear. Probably,
the instance is just related with the fact that a typical distance error
for Tully-Fisher method ($\sim20$\%) corresponds at the Virgo distance to a
linear scale of 3.4~Mpc that comparable with the virial diameter of the
cluster (3.6~Mpc). Large random errors can throw galaxies over the "virial
column" diluting the shape of S-wave infall. Nevertheless, we should
stress that two completely independent sets of observational data on
distances to early-type and late-type galaxies lead to quite compatible
values of $R_0$. As can be seen from Table 1, two samples with Tully-Fisher
distances (E+F) are about 2 times as large in galaxy number as the
samples (A+B+C+D) with precise distances. Because we estimate $R_0$
based on the running median without regard to galaxy weights (distance
errors), the radius $R_0$ for the total sample turns out to be closer to
that for the former samples.

Assuming $R_0$ for the Virgo cluster in the range of
$(5.0 - 7.5)$~Mpc, we obtain from eq.~(4) the total mass of the cluster to be
within $M_T=(2.7 - 8.9)\cdot10^{14}M_{\sun}$. This value is consistent
with the virial mass estimates for the Virgo:
$(5.5\pm1.4)\cdot10^{14}M_{\sun}$ (Hoffman et al. 1980),
$(7.5\pm1.5)\cdot10^{14}M_{\sun}$ (Tully \& Shaya 1984) and
$7.0\cdot10^{14}M_{\sun}$ (Tonry et al. 2000) normalized to the same
Hubble parameter $H_0=72$ km s$^{-1}$ Mpc$^{-1}$. It should be reminded
that this agreement is on very different scales as the zero velocity
radius, $R_0$, is roughly 4 times as large as the virial radius.
This means our results are consistent with there being {\bfseries{}no
additional mass} outside the virial radius of the Virgo cluster.

Two items are worth mentioning in conclusions. The stated method of
estimating $R_0$ value from observational data relays on the assumptions
that a cluster has spherically symmetric shape and that galaxy motions
around the cluster are regular without a significant tangential component.
Deviations from the simple spherical infall scenario have been
considered both theoretically and observationally by many authors
(Peebles 1980, Davis \& Peebles 1983, Lilje et al. 1986, Tonry et al.
2000, Mould et al. 2000, Watkins et al. 2009). In particular, Tonry et
al. (2000) studied a peculiar velocity field around the Local Group and
Virgo cluster in the presence of second attractor (Great Attractor)
situated in Hydra-Centaurus. As seen from their Figs. 20,21, the Great
Attractor with a total mass of $9\cdot10^{15}M_{\sun}$
situated at a distance of 43 Mpc generates significant distortions in
the velocity field over the area of $\sim$1 steradian. Besides, Tully (1988)
found so called effect of "Local Velocity Anomaly" arising by a push of
the Local Void. According to Tully et al. (2009), a bulk motion of
$\sim$260 km~s$^{-1}$ toward the Supergalactic pole may be attributed to
the Local Void. Recently, cosmic velocity flows in the Local Universe
were also studied by Erdogdu et al. (2006), Haugbolle et al. (2007)
and Lavaux (2009).

We would emphasize that observational abilities to determine $R_0$ and
$M_T$ with a better accuracy have not been exhausted yet. With the
derived $R_0$ around 7~Mpc, the zero-velocity surface is situated at a
distance of about 10~Mpc from the observer, i.e. on the Local volume edge.
Hence, one can expect to find some number of the LV galaxies with radial
velocities $\sim(1000 - 2000)$~km~s$^{-1}$, residing in front of Virgo.
A new updated version of CNG (Karachentsev et al. 2010) contains about 40
candidate dwarf galaxies, like VCC1675, with appropriate radial velocities
and rough distance estimates. Their precise distances could be easily
measured via TRGB with facilities of ACS at HST in the one-object-
per-one-orbit mode. The addition of more galaxies with measured distances
in the range $9 - 11$~Mpc from the Local Group would allow a better measurement of
$R_0$, in particular these galaxies could be used to better rule out the
case where there is a large amount of mass outside the virial radius.

{\bf Acknowledgements}. This work was supported by RFBR 07-02-00005,
RFBR-DFG 06-02-04017 and CNRS grants. Authors thank Dmitry Makarov,
Helene Courtois and Stefan Gottloeber for useful discussions.
We would also like to thank the anonimous referees for their valuable
comments which led to significant improvements in the paper.

{}

\newpage
\begin{table}
\caption{Distance datasets for galaxies in/around Virgo.}
\begin{center}
\begin{tabular}{lcrcrrr}
\hline Sample& $\sigma_m$&N(2D)& G(2D)& N(3D)& G(3D) \\ \hline

A:\hspace{1em}TRGB + Ceph & 0.15&  37&  4.1 & 264 &10.8 \\
B:\hspace{1em}SBF (Tonry) & 0.25&  81&  3.6 & 189 & 5.5 \\
C:\hspace{1em}ACSVCS+     & 0.17& 113&  6.2 & 113 & 6.3 \\
D:\hspace{1em}SNIa (Tonry)& 0.10&  13&  3.6 & 22  & 4.7 \\
E:\hspace{1em}TF (IR)     & 0.40&  56&  1.9 & 260 & 4.0 \\
F:\hspace{1em}TF (opt)    & 0.40& 330&  4.5 & 941 & 7.7 \\
\hline
\end{tabular}
\end{center}
\end{table}

\setcounter{table}{1}
\begin{table}
\caption{Radius $R_0$ and its rms error (in Mpc) obtained from bootstrap
realisations.}
\begin{center}
\begin{tabular}{|ccccccc|} \hline

\multicolumn{1}{|c|}{Window}&    
\multicolumn{2}{c|}{Primary distances}&
\multicolumn{2}{c|}{TF distances}&
\multicolumn{2}{c|}{All}\\
\multicolumn{1}{|c|}{(Mpc)}&
\multicolumn{2}{c|}{(samples A+B+C+D) }&
\multicolumn{2}{c|}{(samples E+F) }&
\multicolumn{2}{c|}{samples}\\
\cline{2-7}&
\multicolumn{1}{|c}{Hf}&
\multicolumn{1}{c|}{Vf}&
\multicolumn{1}{c}{Hf}&
\multicolumn{1}{c|}{Vf}&
\multicolumn{1}{c}{Hf}&
\multicolumn{1}{c|}{Vf}\\
\hline
0.8&     6.79&     7.78&     5.07&     5.58&     5.15&     5.82\\
   &$\pm$1.12&$\pm$1.13&$\pm$1.04&$\pm$1.13&$\pm$0.96&$\pm$1.04\\
   &         &         &         &         &         &         \\
1.0&     6.59&     7.54&     4.87&     5.38&     5.00&     5.65\\
   &$\pm$1.05&$\pm$1.03&$\pm$0.88&$\pm$0.97&$\pm$1.00&$\pm$1.07\\
   &         &         &         &         &         &         \\
1.2&     6.46&     7.42&     4.71&     5.24&     4.80&     5.52\\
   &$\pm$1.01&$\pm$0.97&$\pm$0.78&$\pm$0.87&$\pm$0.95&$\pm$1.03\\
\hline
\end{tabular}
\end{center}
\end{table}

\begin{figure}
  \includegraphics[height=\textwidth,keepaspectratio,angle=270]{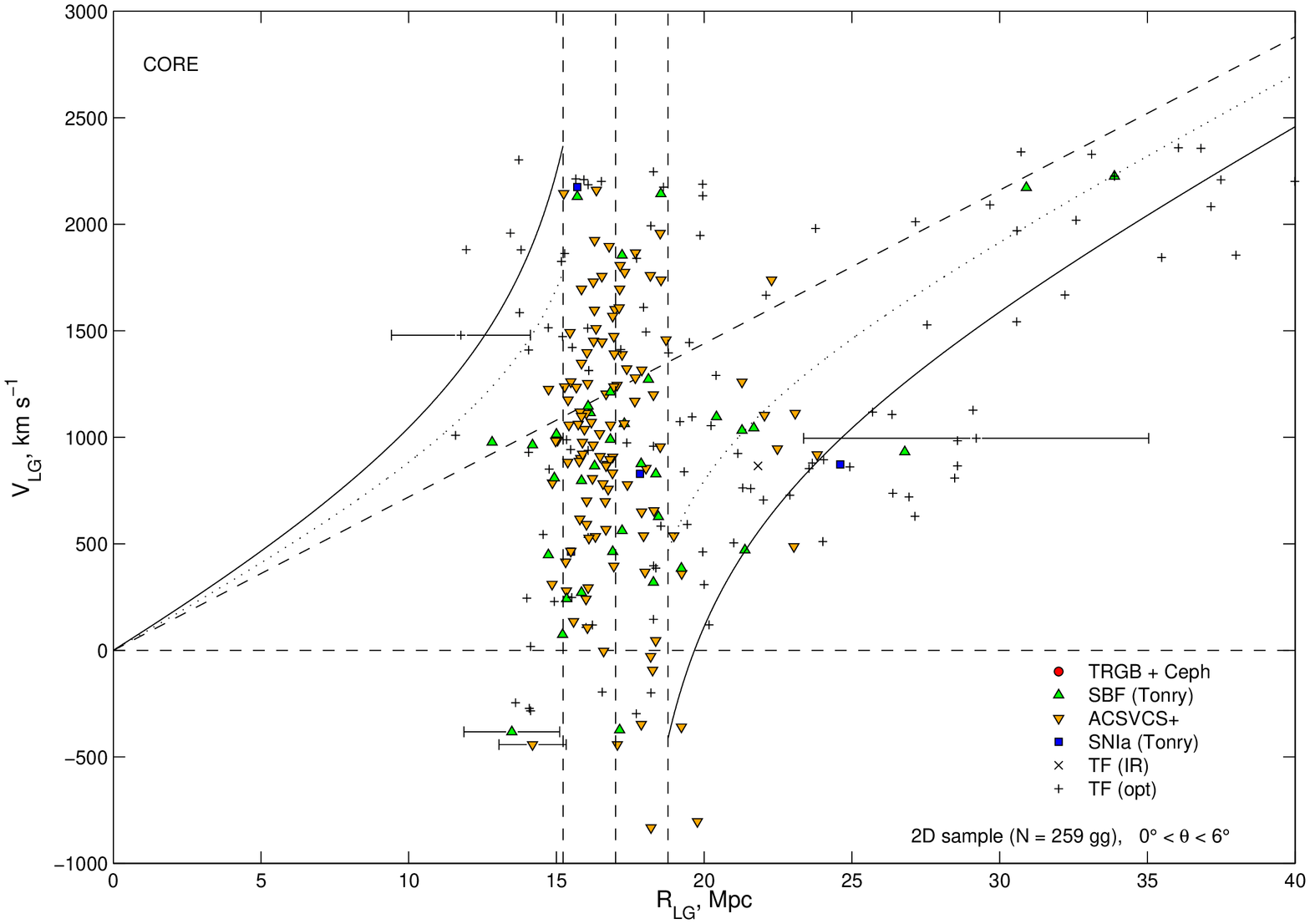}
  \includegraphics[height=\textwidth,keepaspectratio,angle=270]{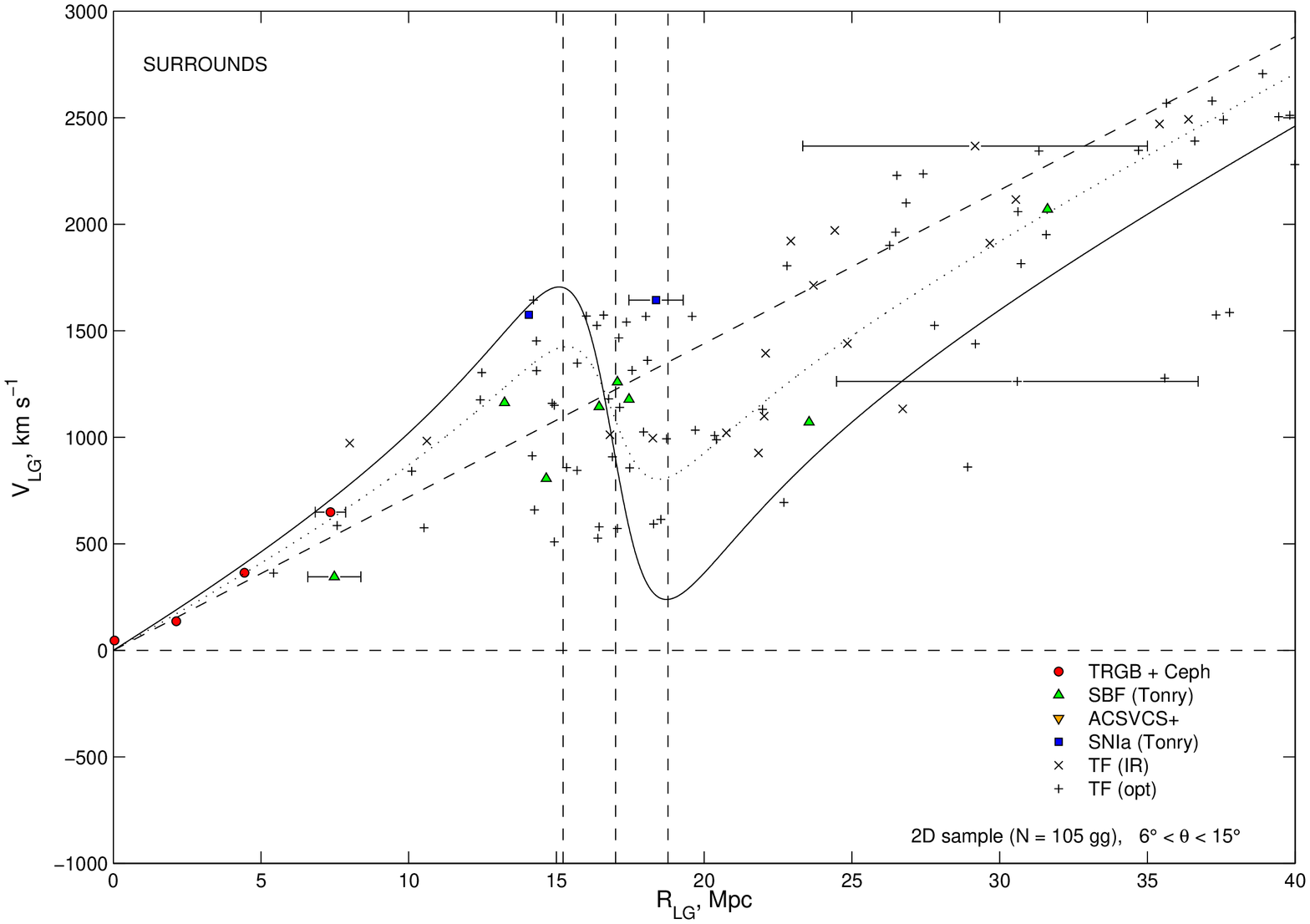}
\end{figure}
\begin{figure}
  \includegraphics[height=\textwidth,keepaspectratio,angle=270]{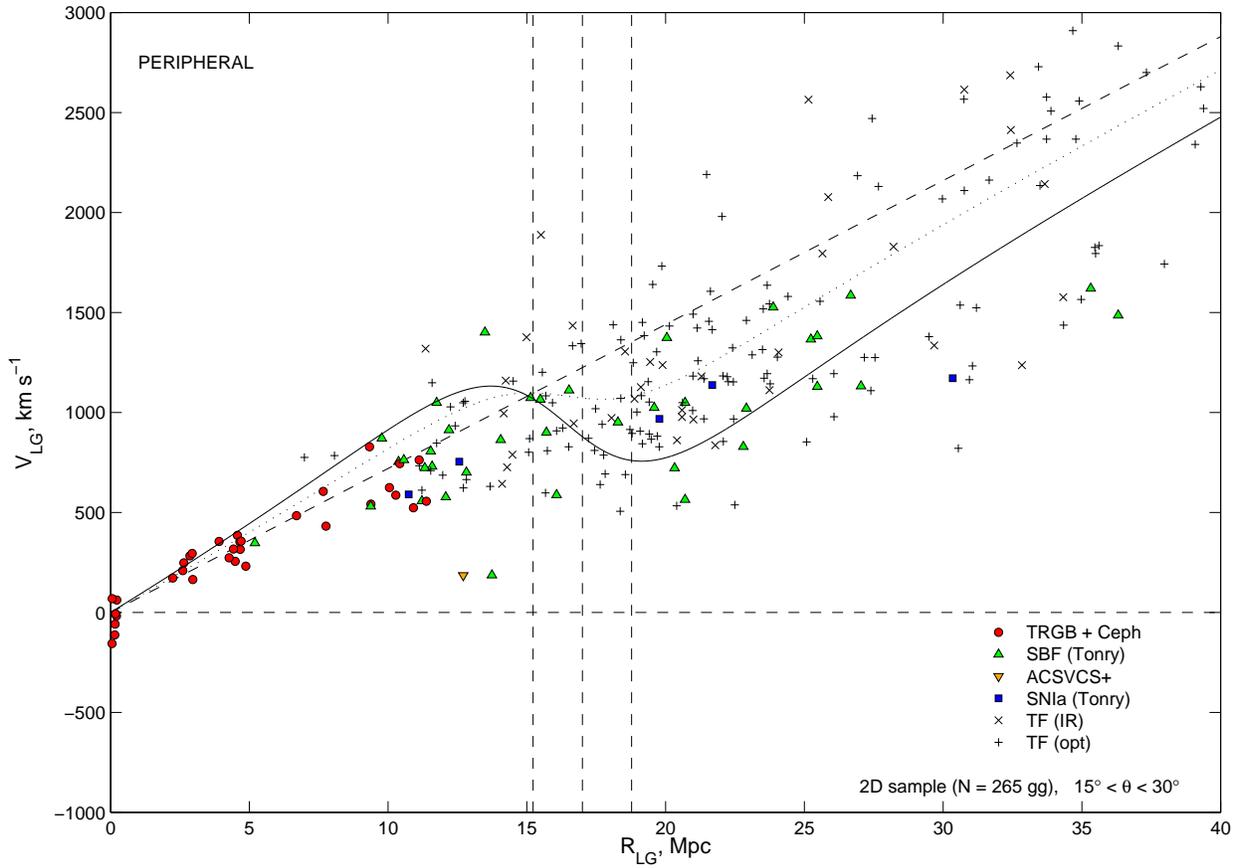}
\caption{The radial velocity vs. distance relation for galaxies in the
  Virgo cluster region with respect to the Local Group centroid. Galaxy
  samples with distances derived by different methods are marked
  by different symbols. The inclined line traces Hubble relation with
  the global Hubble parameter $H_0=72$ km s$^{-1}$ Mpc$^{-1}$ .
  The vertical dashed lines outline the virial zone. Two S-shaped
  lines correspond to a Hubble flow perturbed by virial masses of
  $2.7\cdot10^{14}M_{\sun}$ (dotted) and $8.9\cdot10^{14}M_{\sun}$ (solid)
  as the limiting cases within confidence range.
  {\itshape Top}: the cluster core within angular
  distance $\Theta< 6^{\circ}$,
  both the S-shaped lines indicate the expected infall
  at $\Theta = 0^{\circ}$.
  {\itshape Middle}: wider surroundings with $6^{\circ}<\Theta< 15^{\circ}$,
  the S-shaped lines indicate the infall at $\Theta = 6^{\circ}$.
  The typical distance error bars for each dataset are shown.
  {\itshape Bottom}: peripheric regions with $15^{\circ}<\Theta< 30^{\circ}$,
  the S-shaped lines indicate the infall at $\Theta = 15^{\circ}$.}
\end{figure}

\begin{figure}
  \includegraphics[width=\textwidth,keepaspectratio]{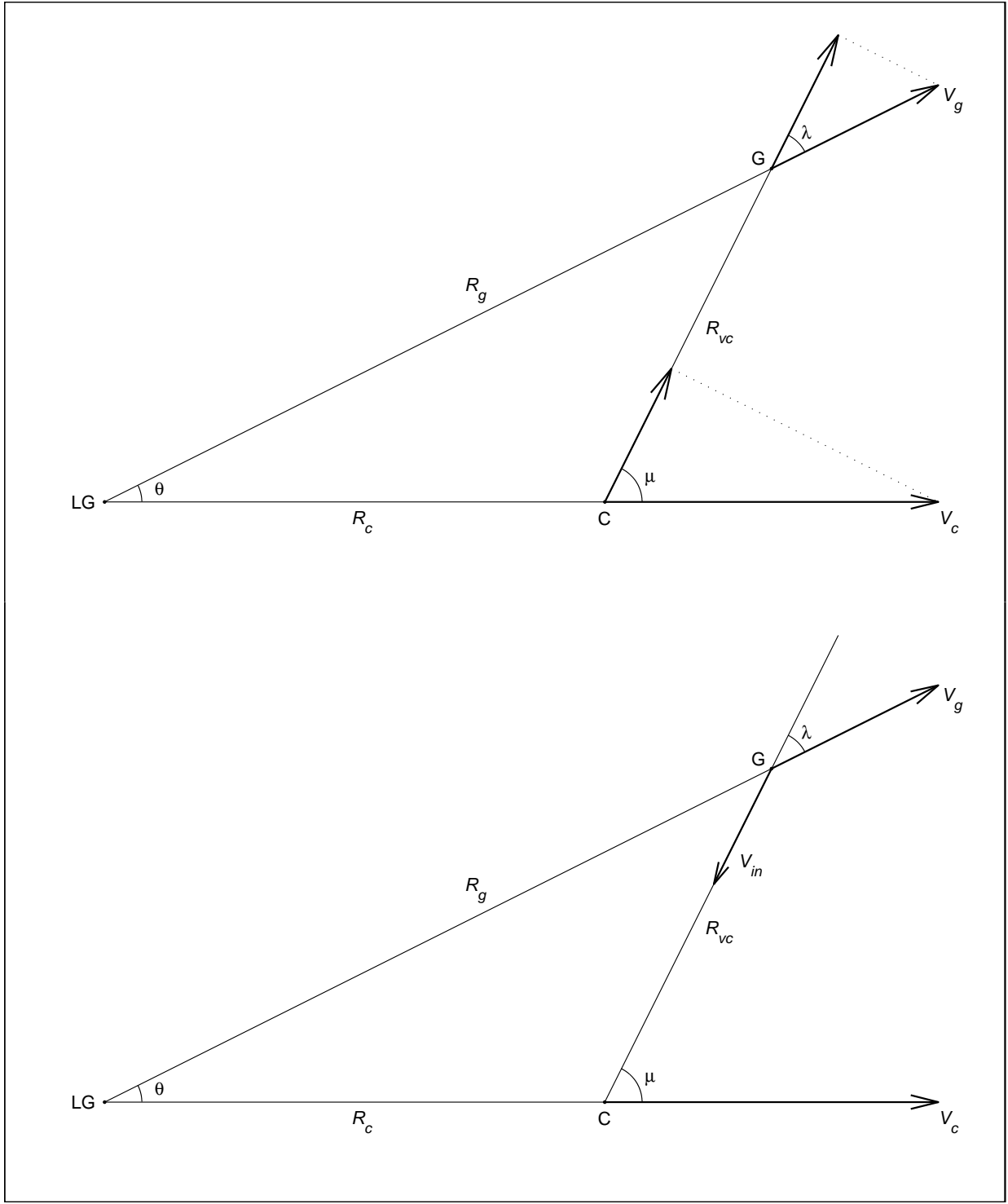}
\caption{Galaxy (G) motion with respect to the cluster center (C) in the
  Local Group (LG) rest frame. {\itshape Top}: in the case of almost
  pure Hubble flow. {\itshape Bottom}: in the case of almost pure
  Virgocentric infall.}
\end{figure}

\begin{figure}
  \includegraphics[height=\textwidth,keepaspectratio,angle=270]{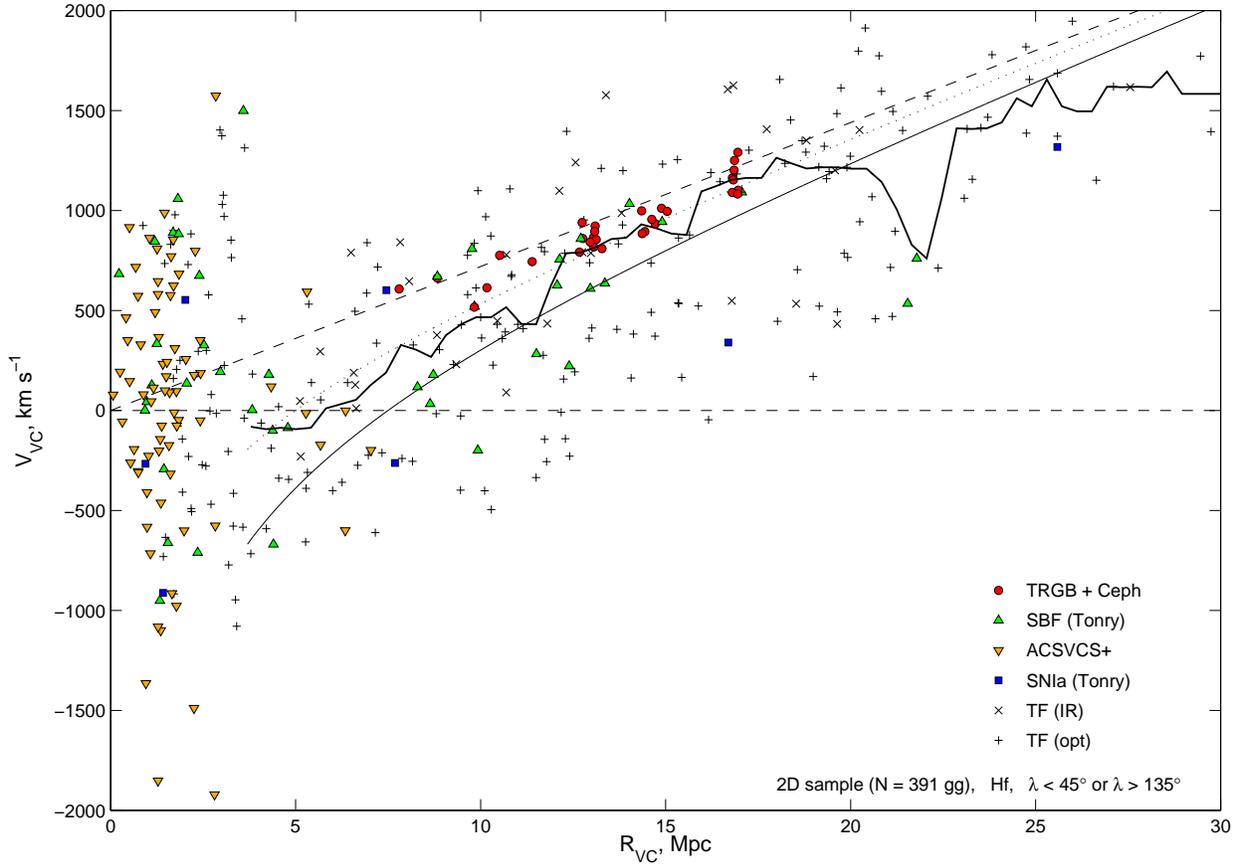}
\caption{Hubble flow around Virgo in the Virgocentric coordinates for 391
  galaxies with $\Theta< 30^{\circ}$ and small angles $\lambda$ regarding
  to the line of sight. The inclined dashed line corresponds to   $H_0=72$
  km~s$^{-1}$ Mpc$^{-1}$, and the polygon curve traces the running
  median with a window of 1 Mpc.
  The dotted and solid curves correspond to a Hubble flow perturbed by a point-like mass of
  $2.7\cdot10^{14}M_{\sun}$ and $8.9\cdot10^{14}M_{\sun}$ respectively as the limiting
  cases.}
\end{figure}

\begin{figure}
  \includegraphics[height=\textwidth,keepaspectratio,angle=270]{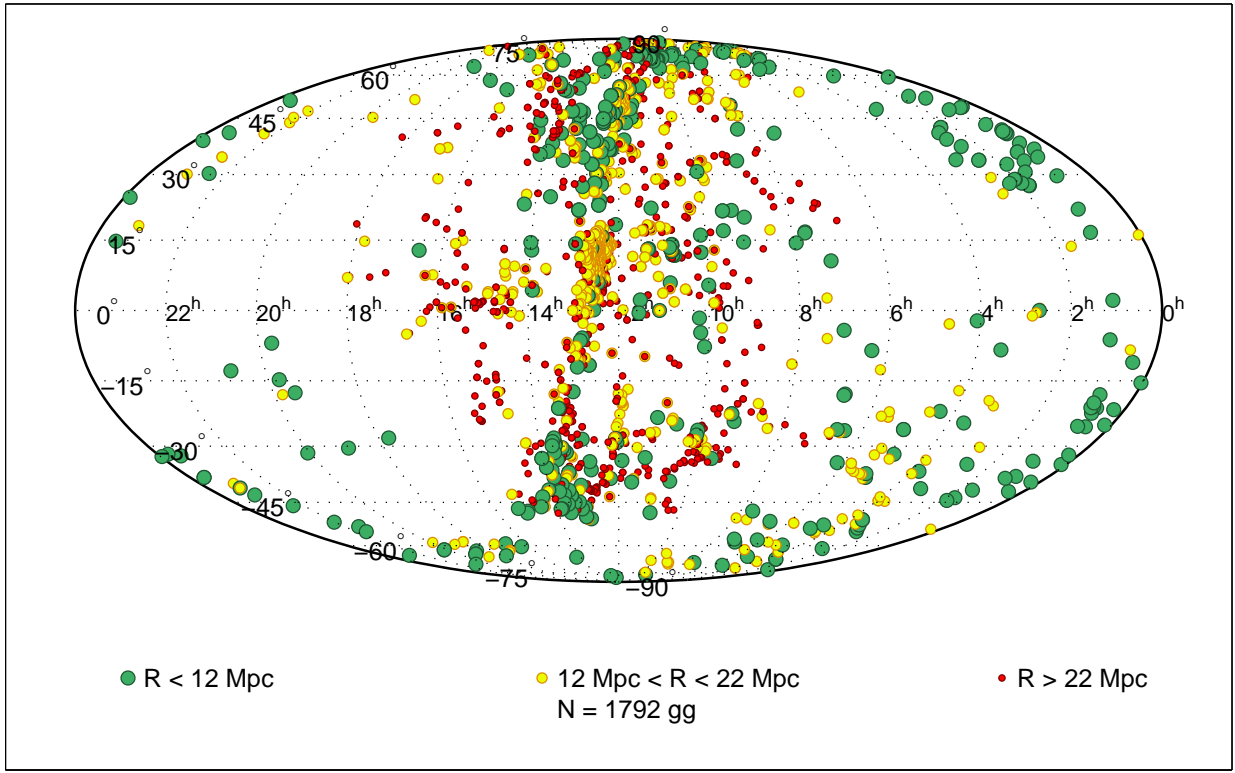}
\caption{Sky distribution of 1792 galaxies with Virgocentric distances
  $R_{vc}<30$ Mpc in equatorial coordinates. Circles of different
  diameters indicate three distance ranges from the LG.}
\end{figure}

\begin{figure}
  \includegraphics[height=\textwidth,keepaspectratio,angle=270]{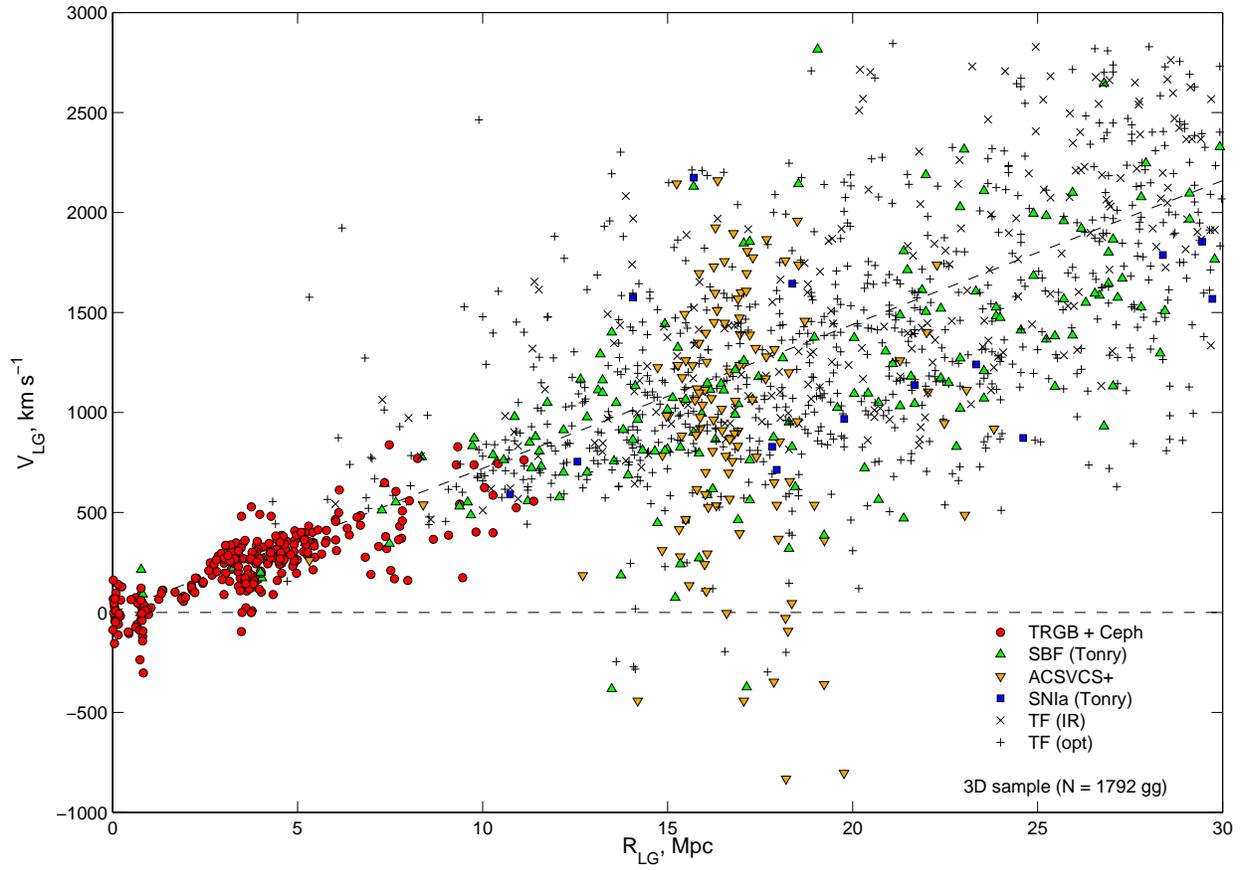}
\caption{The Hubble diagram for 1792 galaxies with Virgocentric distances
  less than 30 Mpc. Symbol indications the same as in Fig.\,1.}
\end{figure}

\begin{figure}
  \includegraphics[height=\textwidth,keepaspectratio,angle=270]{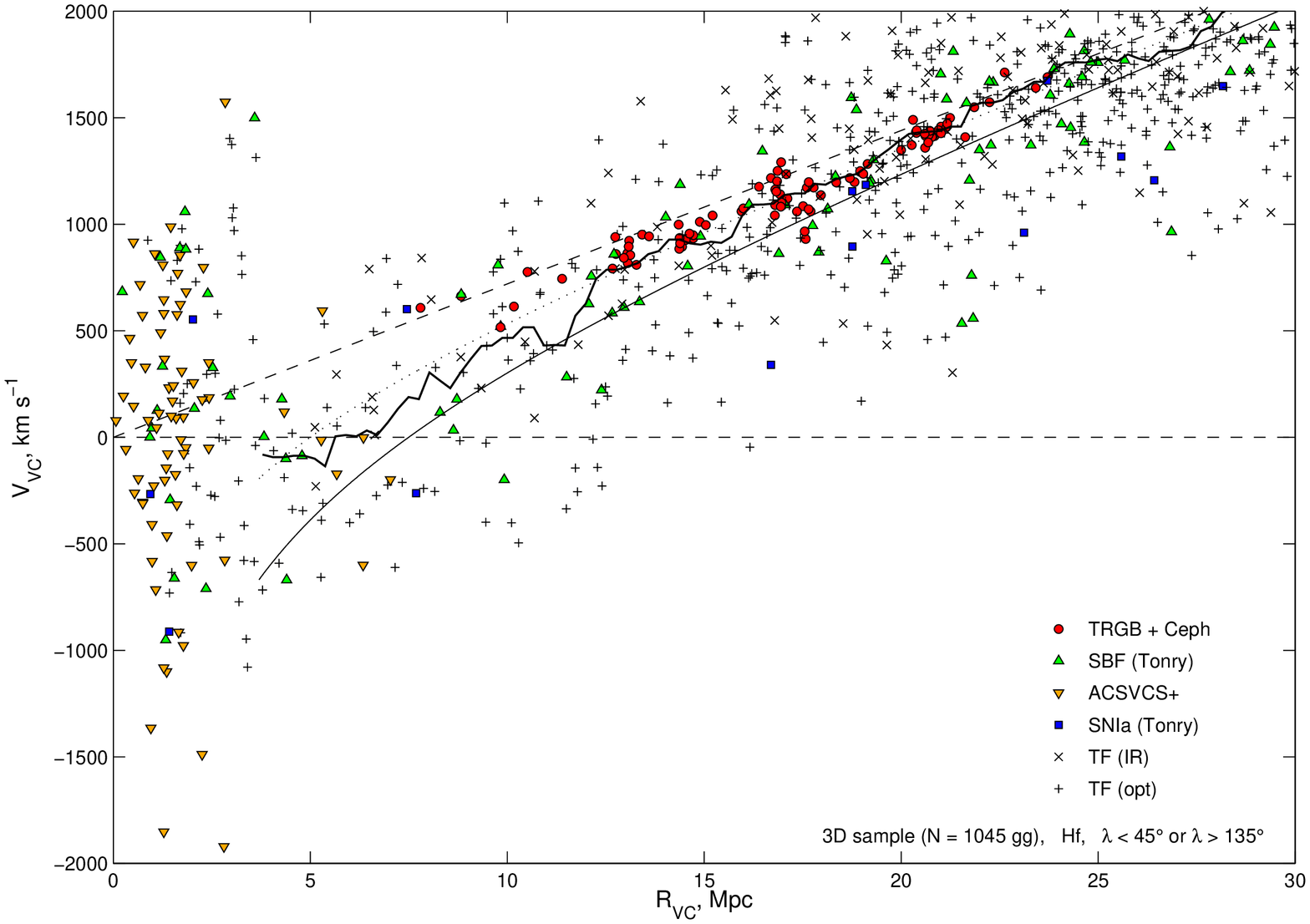}
  \includegraphics[height=\textwidth,keepaspectratio,angle=270]{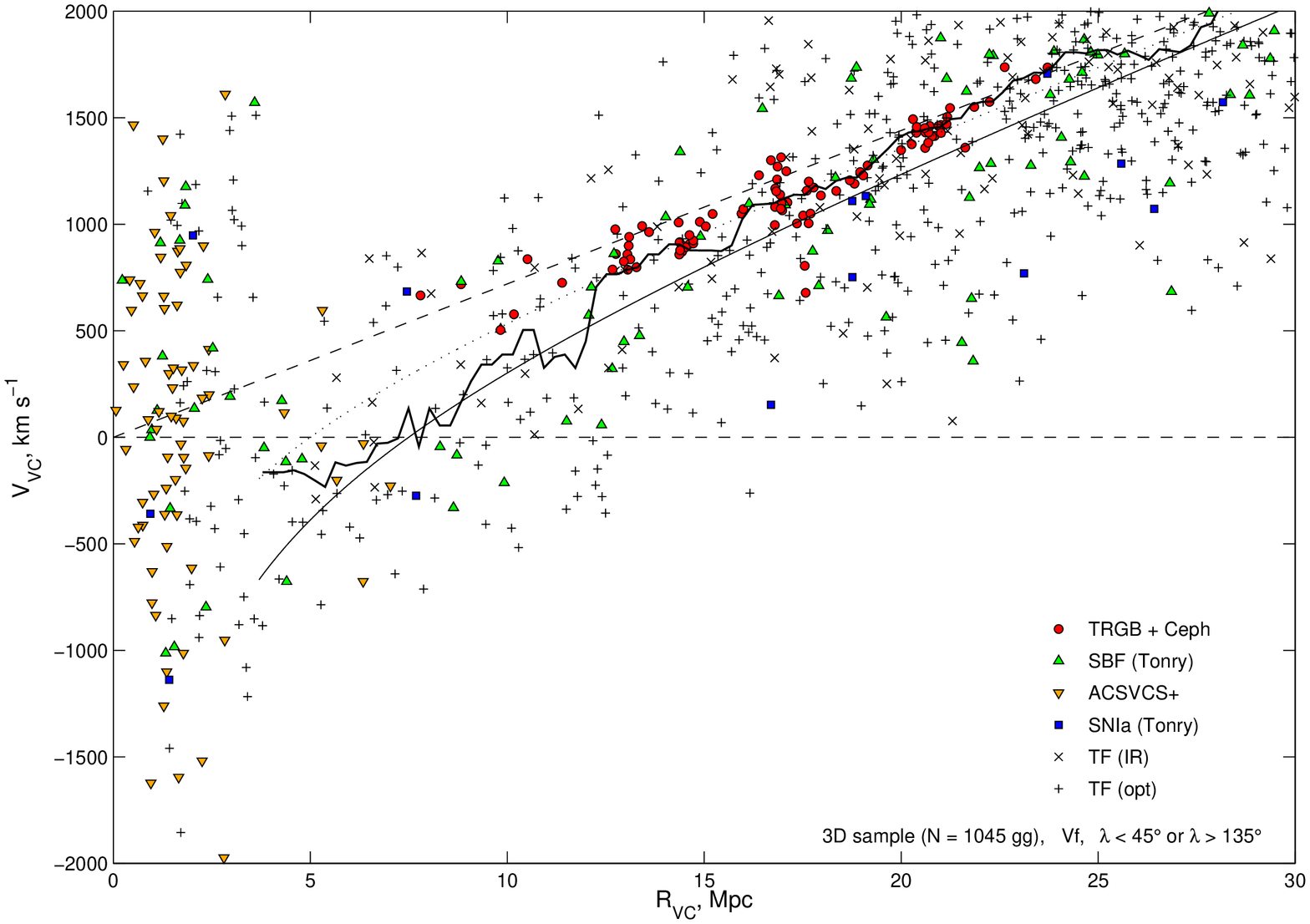}
\caption{The Hubble flow in the Virgocentric coordinates for 1045 galaxies
  with $R_{vc}<30$ Mpc. {\itshape Top}: the case of almost pure Hubble
  flow. {\itshape Bottom}: the case of almost pure Virgocentric infall.
  Only the galaxies with $\lambda<45^{\circ}$ or $\lambda>135^{\circ}$
  are presented. The dotted and solid curves correspond to a Hubble flow perturbed by a point-like mass of
  $2.7\cdot10^{14}M_{\sun}$ and $8.9\cdot10^{14}M_{\sun}$ respectively as the limiting
  cases.}
\end{figure}

\begin{figure}
  \includegraphics[height=\textwidth,keepaspectratio,angle=270]{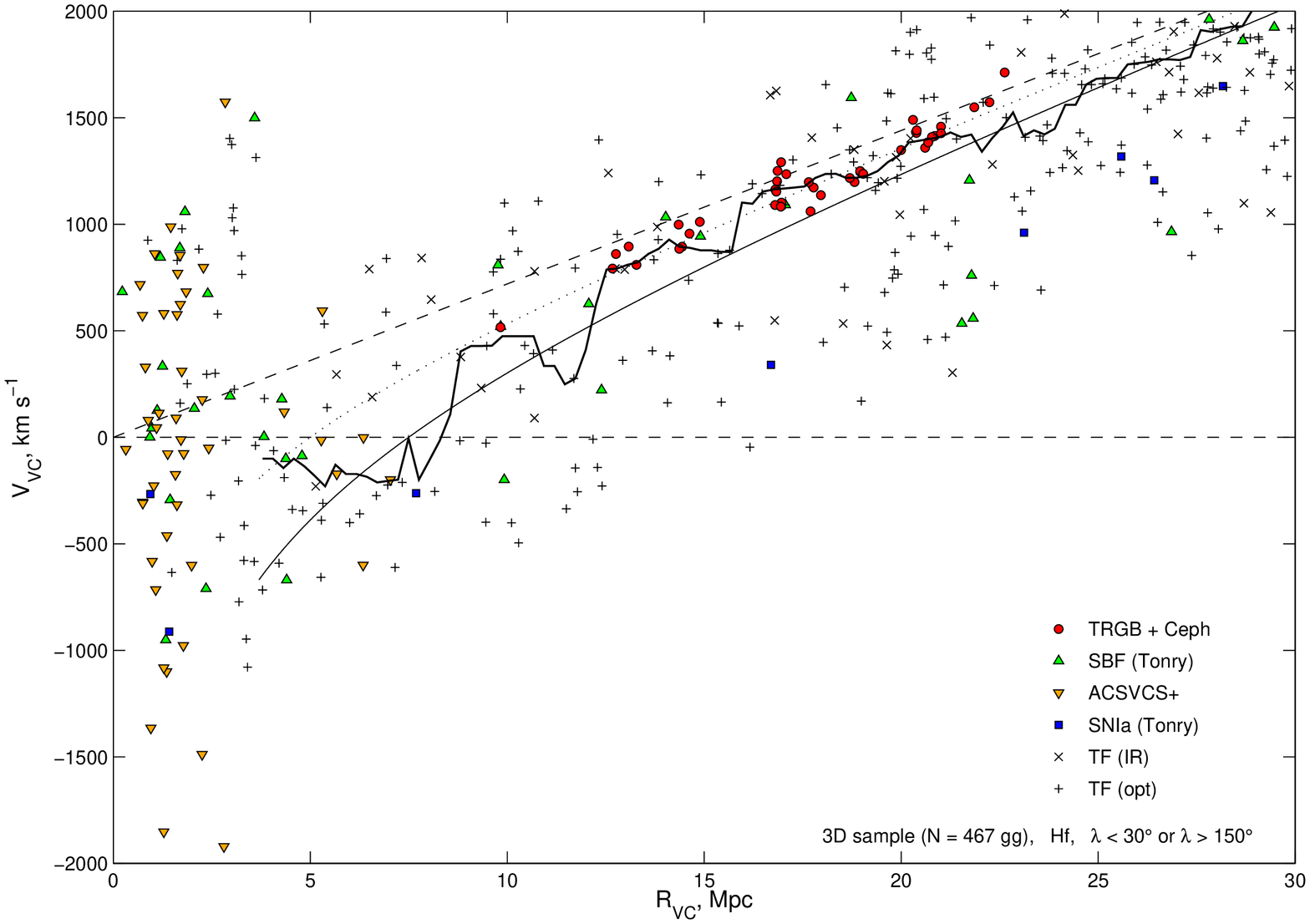}
  \includegraphics[height=\textwidth,keepaspectratio,angle=270]{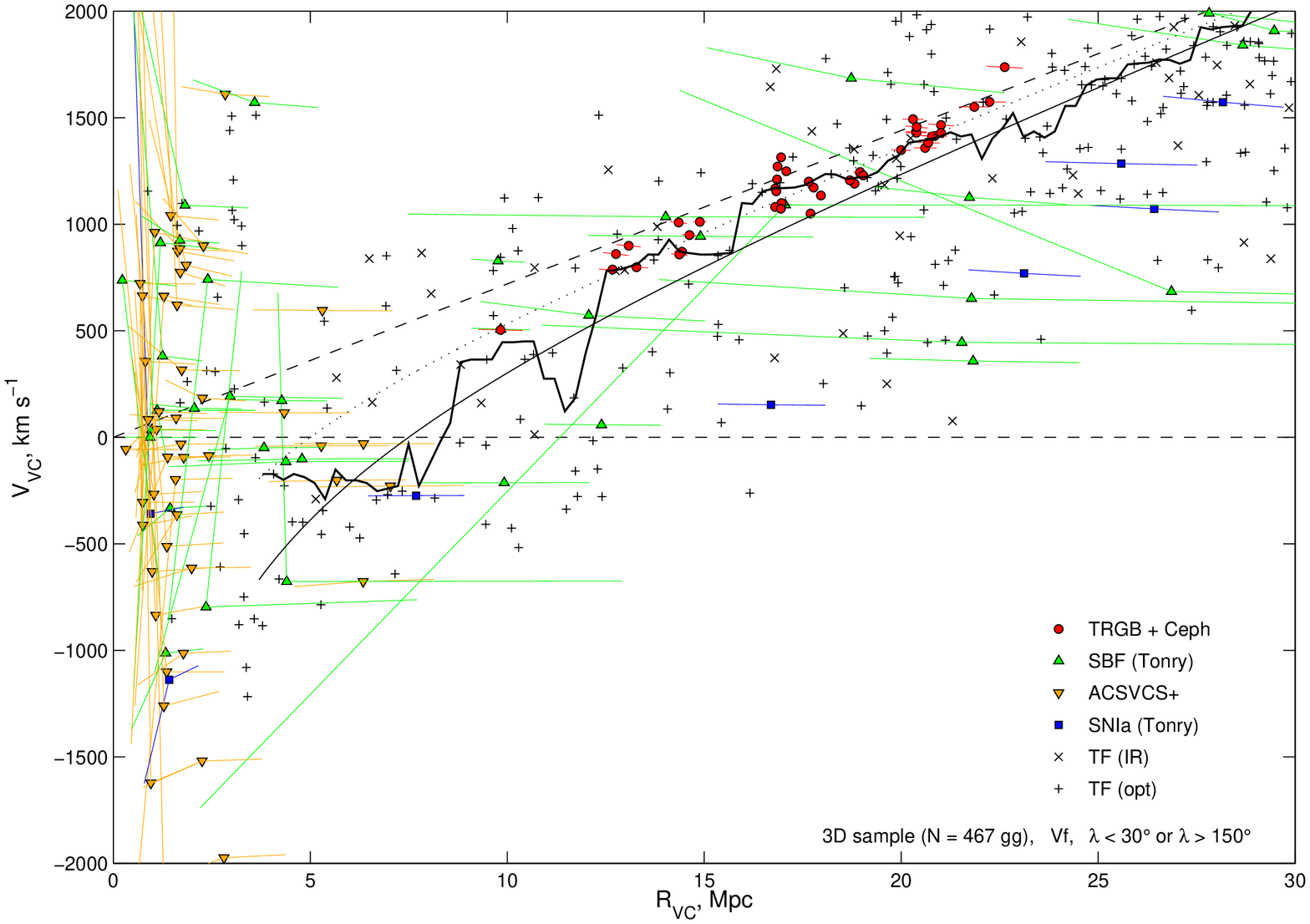}
\caption{The same Hubble diagram as in Fig.\,6, but only for 467 galaxies
  with the angles $\lambda<30^{\circ}$ or $\lambda>150^{\circ}$. In the
  bottom panel all galaxies with precise distances are supplied with
  error bars indicating where the galaxy should be situated if their
  distance from the observer changes by $\pm1\sigma_D$.}
\end{figure}

\end{document}